\RequirePackage{ifpdf}
\ifpdf 
\documentclass[pdftex]{sigma}
\else
\documentclass{sigma}
\fi

\numberwithin{equation}{section}

\newcommand{\half}{\frac{1}{2}}
\newcommand{\RR}{{\mathbb R}}
\newcommand{\R}{{\mathbb R}}
\newcommand{\II}{\mbox{${\mathbb I}$}}
\newcommand{\CC}{{\mathbb C}}
\newcommand{\NN}{{\mathbb N}}
\newcommand{\cP}{{\cal{P}}}

\begin{document}

\allowdisplaybreaks

\renewcommand{\thefootnote}{$\star$}

\renewcommand{\PaperNumber}{090}

\FirstPageHeading

\ShortArticleName{Symmetries of Spin Calogero Models}

\ArticleName{Symmetries of Spin Calogero Models\footnote{This paper is a contribution to the Special
Issue on Dunkl Operators and Related Topics. The full collection
is available at
\href{http://www.emis.de/journals/SIGMA/Dunkl_operators.html}{http://www.emis.de/journals/SIGMA/Dunkl\_{}operators.html}}}

\Author{Vincent CAUDRELIER~$^\dag$ and Nicolas CRAMP\'E~$^\ddag$}

\AuthorNameForHeading{V. Caudrelier and N. Cramp{\'e}}

\Address{$^\dag$~Centre for Mathematical Science, City University,\\
\hphantom{$^\dag$}~Northampton Square, London,
EC1V 0HB,
United Kingdom}
\EmailD{\href{mailto:v.caudrelier@city.ac.uk}{v.caudrelier@city.ac.uk}}

\Address{$^\ddag$~International School for Advanced Studies,
Via Beirut 2-4, 34014 Trieste, Italy}
\EmailD{\href{mailto:crampe@sissa.it}{crampe@sissa.it}}

\ArticleDates{Received September 24, 2008, in f\/inal form December 17,
2008; Published online December 23, 2008}

\Abstract{We investigate the symmetry algebras of integrable spin Calogero systems constructed from Dunkl operators associated to f\/inite Coxeter groups.
Based on two explicit examples, we show that the common view of associating one symmetry algebra to a given Coxeter group $W$ is wrong.
More precisely, the symmetry algebra heavily depends on the representation of $W$ on the spins. We prove this by identifying
two dif\/ferent symmetry algebras for a $B_L$ spin Calogero model and three for $G_2$ spin Calogero model. They are all related to the
half-loop algebra and its twisted versions. Some of the result are extended to any
f\/inite Coxeter group.}


\Classification{70H06; 81R12; 81R50}

\renewcommand{\thefootnote}{\arabic{footnote}}
\setcounter{footnote}{0}

\section{Introduction}

Dunkl operators \cite{Dunkl} were introduced by C.F.~Dunkl as part of a program on polynomials in several variables with ref\/lection symmetries related to
f\/inite ref\/lection groups (or equivalently f\/inite Coxeter groups). Besides this important mathematical motivation, they quickly became fundamental objects
in the study of integrable quantum mechanical many-body systems introduced by F.~Calogero and B.~Sutherland~\cite{Calo,Suth} and generalized to any Weyl group
in~\cite{PO}. Indeed they can be viewed as generalizations of momentum operators with which they share the crucial property of forming an Abelian algebra.
This fact allows one to implement the exchange operator formalism described in \cite{Poly} producing many-body integrable Hamiltonians. In \cite{Poly2,HW},
it was realized that this could be pushed further by representing the Coxeter group on spins\footnote{Here the spin is to be understood as an internal
degree of freedom taking $N$ dif\/ferent values.}. This gave $L$-particle integrable spin Calogero systems related to the group~$A_L$. Then one could play
several games such as changing the type of potential and/or the underlying group while maintaining integrability.

Another big step was taken in \cite{BGHP} where the ``FRT formalism'' \cite{FRT},
so powerful with spin chain type integrable models, was used in connection with the algebraic properties of the Dunkl operators to construct the symmetry
algebra. By this, we mean the algebra of the operators commuting with the Hamiltonian of the system (and its hierarchy). In the case of~\cite{BGHP}, it
is a non Abelian inf\/inite dimensional algebra which contains the whole hierarchy of the integrable~$A_L$ spin Sutherland model in its centre: the Yangian of $\mathfrak{gl}_N$ \cite{Dri}.
This type of symmetry algebra is a very ef\/f\/icient tool to study and, in particular, to compute explicitly the spectrum, the eigenvectors and the correlation functions of a large class of integrable models:
Sutherland model~\cite{TU,Tak,Ug}, Haldane--Shastry spin model \cite{HH,HHTBP,BPS,Hik}, non-linear Schr\"odinger equation~\cite{MW,RS} and Hubbard model~\cite{UK}.

The method developed in \cite{BGHP} was then derived in various ways, again accommodating various groups and potentials, to identify the corresponding symmetry algebras. From there, following several results in the literature, some standard views imposed themselves as granted: for instance, the symmetry
algebra of an $A_L$ Sutherland type model \cite{BGHP} should be a Yangian while the Calogero counterpart (without harmonic conf\/inement)
should be the corresponding half-loop algebra. Similarly, for $B_L$ Sutherland models, the ref\/lection algebra \cite{Skly} should be symmetry algebra
\cite{caduk} with the corresponding twisted half-loop algebra as symmetry algebra of the Calogero counterpart (this paper).

Actually, we show in this paper that this
is not true and in general, it is not enough to specify the type of potential and the underlying Coxeter group to identify the correct symmetry algebra of the
associated integrable model. In addition, one has to give the spin representation of the Coxeter group.
This is examplif\/ied in this paper by working with rational Dunkl operators (hence producing Calogero type potentials) based on two particular
f\/inite Coxeter groups: $B_L$~and~$G_2$. In the f\/irst three sections, we recall the background on f\/inite ref\/lection groups, half-loop algebras and the link to
dynamical spin integrable systems that is necessary for our purposes. In Section~\ref{reflection}, we deal with $B_L$ spin Calogero model for which we identify
two dif\/ferent symmetry algebras by choosing two dif\/ferent ways of representing $B_L$ on the spin conf\/iguration space. In Section~\ref{G2}, we apply the same
strategy with three dif\/ferent spin representations of $G_2$ to obtain three dif\/ferent symmetry algebras. All the symmetry algebras are related to the
$\mathfrak{gl}_N$ half-loop algebra and its twisted versions. Various generalizations of our discussion are collected in Section~\ref{general}. Our
conclusions are gathered in the last section.

\section[Dunkl operators associated to finite reflection groups]{Dunkl operators associated to f\/inite ref\/lection groups}

Here we recall the ingredients we need from the theory of f\/inite ref\/lection groups (see e.g.~\cite{hum}) and Dunkl operators~\cite{Dunkl}.
A ref\/lection in the real Euclidean space $\RR^L$ endowed with the scalar product $\displaystyle(e_i,e_j)=\delta_{ij}$, where $\{e_i\}_{i=1,\dots,L}$ are the
canonical basis vectors of $\RR^L$, is a linear opera\-tor~$s_\alpha$ ($\alpha\in \RR^L, \alpha\neq 0$) on $\RR^L$ def\/ined by
\begin{gather*}
\forall\,\mu\in \RR^L,\qquad s_\alpha (\mu)=\mu-2 \frac{(\mu,\alpha)}{(\alpha,\alpha)}\alpha .
\end{gather*}
It sends the vector $\alpha$ to $-\alpha$ and leaves invariant the hyperplane $H_\alpha$ orthogonal to $\alpha$.
A root system $\Phi$ is a subset of $\RR^L$ satisfying, for all $\alpha\in \Phi$,
\begin{gather*}
 (\lambda\in\RR,\ \alpha\in\Phi\quad \text{and}\quad\lambda\alpha\in\Phi)\ \Rightarrow \ \lambda=\pm 1 ,\\
 s_\alpha \Phi = \Phi .
\end{gather*}
$\Phi$ is a f\/inite set and the group generated by $\{s_\alpha\ |\ \alpha\in \Phi\}$ is a f\/inite
subgroup $W$ of $O(\RR^L)$ called the ref\/lection group associated to $\Phi$.
Any root system can be written as $\Phi=\Phi^+\cup (-\Phi^+)$ ($\Phi^+$ is called the positive root system).
A simple system $\Delta$ is a subset of $\Phi$ such that $\Delta$ is a vector space basis of ${\rm span}_\mathbb{R} \Phi$ and
each $\alpha\in \Phi$ is a linear combination of $\Delta$ with coef\/f\/icients all of the same sign.

An important result is that the ref\/lection group $W$ is generated by $\{s_\alpha\ |\ \alpha\in \Delta\}$ subject only
to the relations, for $\alpha, \beta\in \Delta$
\begin{gather*}
(s_\alpha s_\beta)^{m(\alpha,\beta)}=1
\end{gather*}
with $m(\alpha,\alpha)=1$ and $m(\alpha,\beta)>1$ ($\alpha \neq \beta$). That is to say, by def\/inition, $W$ is a Coxeter group.
The possible values of $m(\alpha,\beta)$ such that the group $W$ is f\/inite provides
a classif\/ication of the f\/inite ref\/lection groups.

To each Coxeter group $W$, one can associate Dunkl operators which are dif\/ferential operators acting on functions $\varphi:\RR^L\rightarrow \mathbb{C}$.
Let us def\/ine the natural action of $W$ on such functions by
\begin{gather*}
(\hat s \varphi)(\mu)=\varphi(s^{-1}(\mu))
\end{gather*}
with $s\in W$ and $\mu\in \RR^L$. The map $s\mapsto \hat s$ is a representation
of the group $W$.
A function $k:\Phi\rightarrow \mathbb{C}$ is called a multiplicity function if
\begin{gather*}
\forall\,\alpha,\beta\in \Phi, \qquad k(\beta)=k(s_\alpha(\beta)).
\end{gather*}
Let $k$ be given. For $\xi\in \R^L$, the Dunkl operator $d_\xi$ acts on $C^1(\RR^L)$ and is def\/ined by
\begin{gather*}
d_\xi = -i\partial_{\xi}+i\sum_{\alpha\in \Phi^+}
k(\alpha)\frac{(\alpha,\xi)}{(\alpha,x)}\hat s_\alpha ,
\end{gather*}
where $\partial_{\xi}$ is the derivative in the direction $\xi$,
$x=(x_1,\dots,x_L)^t$ with $x_i$ the operator multiplication by $x_i$. Note that our choice of Dunkl operators yields hermitian operators provided $k$ is
real-valued. This is important for quantum mechanical applications.
The Dunkl operators do not depend on the choice of the positive
root system $\Phi^+$ and have the following fundamental properties\footnote{Strictly speaking, the proof in \cite{Dunkl} is given for slightly
dif\/ferent Dunkl operators simply related to ours by the gauge transformation $d_\xi\mapsto \phi^{-1}(x)d_\xi\phi(x)$ where $\phi$ is a
$W$-invariant function $\phi(s(x))=\phi(x)$ for all $s\in W$.}
\begin{proposition}[\cite{Dunkl}]
The Dunkl operators are $W$-equivariant, i.e.\ for any $s\in W$, we get
\begin{gather*}
\hat sd_\xi \hat s^{-1}= d_{s(\xi)}.
\end{gather*}
The Dunkl operators commute, i.e.\ for any $\xi,\zeta\in \R^L$, we get
\begin{gather*}
d_{\xi}d_\zeta=d_\zeta d_{\xi}.
\end{gather*}
\end{proposition}

\section{Half-loop algebra and twists}

This section is meant to give necessary def\/initions and notations to handle algebraic structures related to some integrable models.

The $\mathfrak{gl}_N$ half-loop algebra\footnote{Also sometimes called Gaudin algebra.} is the unital algebra $\mathfrak{gl}_N[z]$ of polynomials in an indeterminate $z$ with
coef\/f\/icients in $\mathfrak{gl}_N$: it is generated by $\{e_{ij}\,z^n\ | \ 1\leq i,j\leq N, n=0,1,2,\dots\}$
where $e_{ij}$ are the generators of $\mathfrak{gl}_N$ satisfying the commutation relations
\begin{gather*}
\left[e_{ij},e_{k\ell}\right]=\delta_{jk}e_{i\ell}-\delta_{i\ell}e_{kj} .
\end{gather*}

Let $E_{ij}$ be the matrix with $1$ on the entry $(i,j)$ and $0$ elsewhere.
The map $e_{ij}\mapsto E_{ij}$ provides a representation of $\mathfrak{gl}_N$. In the following, we will use auxiliary space notations to simplify computations.
We def\/ine the monodromy matrix of the half-loop algebra as follows
\begin{gather*}
T(u)=\sum_{i,j=1}^N E_{ij}\otimes\sum_{n=0}^\infty\frac{e_{ji}z^{n}}{u^{n+1}}=\sum_{i,j=1}^N \frac{E_{ij}\otimes e_{ji}}{u-z}
\in {\rm End}(\CC^N)\otimes\mathfrak{gl}_N[z][[u^{-1}]] ,
\end{gather*}
where $u$ is a formal parameter called the spectral parameter and ${\rm End}(\CC^N)$ is the auxiliary space used to pack nicely the generators. The def\/ining relations of the half-loop algebra can be written
\begin{gather}
\label{rtt}
\left[T_a(u),T_b(v)\right]=\left[T_a(u)+T_b(v),r_{ab}(u-v)\right],
\end{gather}
where $a$ and $b$ denote two copies of the auxiliary space ${\rm End}(\CC^N)$,
$T_a(u)=T(u)\otimes \II_N$, $T_b(u)= \II_N\otimes T(u)$, $\II_N$ is the identity in ${\rm End}(\CC^N)$,
 and
\begin{gather*}
r_{ab}(u)=\sum_{i,j=1}^N\frac{E_{ij}\otimes E_{ji}}{u}\equiv \frac{P_{ab}}{u}
\end{gather*}
is a classical $r$-matrix solution of the classical Yang--Baxter equation \cite{BD}.

Let $\sigma\in {\rm End}(\CC^N)$ with $\sigma^n=\II_N$ for some $n\in\NN$. The
eigenvalues of $\sigma$ are the powers of the $n$-th root
of unity $\tau=e^{\frac{2i\pi}{n}}$. We denote $A$ the inner automorphism
of $\mathfrak{gl}_N$
associated to $\sigma$. Its action on the generators of $\mathfrak{gl}_N$
is given by
\begin{gather*}
A: \ \ e_{ij}\mapsto (\sigma)_{jn}\,e_{mn}\,(\sigma^{-1})_{mi} .
\end{gather*}
We extend this action to $\mathfrak{gl}_N[z]$ in the natural way and we def\/ine
\begin{gather*}
\mathfrak{gl}_N[z]^\sigma=\{M(z)\in\mathfrak{gl}_N[z]\, | \, A\,M(z)=M(\tau z)\}.
\end{gather*}
One can see that $\mathfrak{gl}_N[z]^\sigma$ is a subalgebra of $\mathfrak{gl}_N[z]$ which is
called \textit{the (inner) twisted half-loop algebra of order $n$}.
Now we proceed to derive the commutation relations of this algebra. For $k\in\NN$, we introduce the projectors
\begin{gather*}
\cP_k=\frac{1}{n}\sum_{j=0}^{n-1}\tau^{-jk}A^j.
\end{gather*}
Note that there are only $n$ such projectors as $\cP_{n+k}=\cP_k$.
This allows us to def\/ine a surjective projection map from $\mathfrak{gl}_N[z]$ to $\mathfrak{gl}_N[z]^\sigma$ by $e_{ij}z^k\mapsto \cP_k e_{ij}z^k$.
In turn, this maps $T(u)$ to
\begin{gather}
\label{maps}
B(u)=\frac{1}{n}\sum_{j=0}^{n-1}\tau^j\,\sigma^j\,T(\tau^j u)\sigma^{-j}
\in {\rm End}(\CC^N)\otimes\mathfrak{gl}_N[z]^\sigma[[u^{-1}]],
\end{gather}
which contains the generators of $\mathfrak{gl}_N[z]^\sigma$ by construction.
Next we have \cite{NY}
\begin{proposition}
$B(u)$ satisfies the symmetry property
\begin{gather*}
\forall\,j=0,\dots,n-1, \qquad B(u)=\tau^j\sigma^jB(\tau^j u)\sigma^{-j},
\end{gather*}
and encodes the commutation relations of $\mathfrak{gl}_N[z]^\sigma$ as
\begin{gather}
\label{twisted_rel}
\left[B_a(u),B_b(v)\right]=\frac{1}{n}\sum_{j=0}^{n-1}\left[\tau^j\, B_a(u)+B_b(v),
\frac{(\sigma^{j})_bP_{ab}(\sigma^{-j})_b}{u-\tau^j\,v}\right].
\end{gather}
\end{proposition}
For our purposes, the following facts are important. If $b(u)={\rm Tr}\,B(u)$, where the trace is taken over the
auxiliary space ${\rm End}(\CC^N)$, then
it follows from (\ref{twisted_rel}) that
\begin{gather}
\label{prop}
\left[b(u),B(v)\right]=0,\qquad \left[b(u),b(v)\right]=0.
\end{gather}

\section{Link with integrable dynamical spin models}\label{sec:link}

The Dunkl operators allow us to construct integrable Hamiltonians by implementing the strategy discussed in \cite{Poly}. Indeed, any polynomial
in the Dunkl operators commutes with the independent and
two by two commuting operators $\{d_{e_i}\,|\,i=1,\dots,L\}$.
When it is of order~$2$, this polynomial can usually be regarded as an Hamiltonian which is then integrable.

The next step is to consider Hamiltonians for particles with $N$ internal degrees of freedom. The construction to introduce these ``spins''
has been pioneered in \cite{Poly2,HW} and the symmetry of the corresponding models has been shown in~\cite{BGHP}.
The point is that the Dunkl operators only act on the positions $x_i$ while now
the wavefunctions become $\varphi(x|s)$, $x\in \RR^L$, $s\in (\CC^N)^{\otimes M}$.
Let us remark that $M$ may be dif\/ferent from $L$ in general. So one has to come up with a method allowing the Dunkl operators to act on the spins while maintaining
their nice properties. This is done with a suitable projector
\begin{gather*}
\Lambda=\frac{1}{|W|}\sum_{w\in W}\hat{w}R_w.
\end{gather*}
Note that we require that the wavefunctions satisfy the following generalized statistics\footnote{Called this way because in the case
$W=A_L$ and with the permutation representation on positions and spins, these conditions amount to consider bosonic wavefunctions.}
 for $w\in W$,
\begin{gather*}
\hat w R_w \varphi(x|s)=\varphi(x|s),
\end{gather*}
where $R: W\to {\rm End}((\CC^N)^{\otimes M})$ is a representation of~$W$.

One then acts with $d_\xi\Lambda$ on the wavefunctions.
Unfortunately, usually $[d_\xi\Lambda,d_\zeta\Lambda]\neq 0$ so the idea to use directly polynomials in $d_\xi\Lambda$ to get an integrable hierarchy fails.
The idea of~\cite{BGHP} is to use the monodromy matrix formalism and the underlying algebraic structures to circumvent this problem.
In our context, consider a set of M vectors $\{\xi_1,\dots,\xi_M\}$. Then, the matrix
\begin{gather*}
T_a(u)=\sum_{k=1}^M\frac{P_{ak}}{u-d_{\xi_k}}
\end{gather*}
provides a representation of the half-loop algebra (\ref{rtt}).
At this point, let us emphasize that the auxiliary space has no `physical' meaning (in the sense it does not act on the wavefunctions) but it is a
powerful and convenient tool to deal with the inf\/inite algebras we are interested in.
Then, the `physical' operators are contained in the entries of the matrix.

One obtains
$B(u)$ through (\ref{maps}) with the properties (\ref{prop}).
The crucial part is to show that, for a suitable choice of $\xi$ and of the twist,
$B(u)\Lambda$ is also a representation of the algebra (\ref{twisted_rel}).
To prove this statement, it is suf\/f\/icient to show that
$\Lambda B(u)\Lambda=B(u)\Lambda$ which is guaranteed by, for $g$ any generator of $W$,
\begin{gather}
\label{commute_lambda}
\hat g R_g B(u) \Lambda=B(u)\Lambda.
\end{gather}
Finally, using properties (\ref{prop}) for $B(u)\Lambda$, we conclude that
${\rm Tr}\, B(u)\Lambda$ provides a hierarchy of commuting operators when expanding in powers of $u^{-1}$.
To prove the integrability, it is now suf\/f\/icient to prove that in this hierarchy there
are $L$ independent quantities.
Noting that these operators are polynomials in $d_\xi$ times $\Lambda$, we may choose one of them as the Hamiltonian acting both on the position
and spin degrees of freedom of the wavefunctions.
As a by-product, but not the least,
we prove that this Hamiltonian has the twisted half-loop algebra for symmetry algebra.

It is very important to realize that it is the choice of $R$, the representation of $W$ on the spins, together with the requirement
(\ref{commute_lambda}), that imposes the form of $B(u)$ and hence the symmetry algebra. This is the essential message of this paper.

In the rest of this paper, we detail this construction
for two Coxeter groups, $B_L$ and $G_2$. In particular, we get the surprising
result that spin Calogero-type integrable models built from Dunkl operators
associated to the same Coxeter group can have dif\/ferent symmetry algebras.

\section[Reflection group $B_L$]{Ref\/lection group $\boldsymbol{B_L}$}\label{reflection}

We consider the Coxeter group $B_L$. It is
generated by $L$ elements $t_1,\dots,t_{L-1}$ (generating the
Coxeter group $A_{L-1}$) and $r$ with def\/ining relations
\begin{gather*}
 r^2=1 ,\qquad t_i^2=1 ,\qquad i=1,\dots,L-1 ,\\
 (t_it_{i+1})^3=1 ,\qquad i=1,\dots,L-2 ,\qquad (t_it_j)^2=1 ,\qquad |i-j|>1,\\
 (rt_{L-1})^4=1 ,\qquad (rt_j)^2=1 ,\qquad j\neq L-1 .
\end{gather*}

\subsection[The usual $B_L$ spin Calogero model]{The usual $\boldsymbol{B_L}$ spin Calogero model} \label{sec:1}

To f\/ix ideas, we start with a known model \cite{Ya}.
We choose the usual positive root system $\Phi^+=\{e_i\pm e_j,e_k\ |\ 1\leq i<j\leq
L,1\leq k\leq L\}$.
Then, the representation of the generators of $B_L$ associated to
simple roots are
\begin{gather*}
t_i=s_{e_i-e_{i+1}}=\left(\begin{array}{c c c c}
\II_{i-1} \\
 & 0 & 1 &  \\
 & 1 & 0 &   \\
 &   &   & \II_{L-i-1}
\end{array}\right)
\qquad\text{and}\qquad
r=s_{e_L}=\left(\begin{array}{c c}
\II_{L-1}&\\
& -1
\end{array}\right),
\end{gather*}
where $\II_{j}$ is the $j\times j$ unit matrix. In this case, the Dunkl operators are given by
\begin{gather}
\label{def2_dunkl}
d_{e_k}\equiv d_k=-i\partial_{e_k}+ik_l\sum_{j \neq k}\left(\frac{1}{x_k-x_j}\hat s_{e_k-e_j}+\frac{1}{x_k+x_j}\hat s_{e_k+e_j}\right)+\frac{ik_s}{x_k}\hat s_{e_k}.
\end{gather}
The constants $k_l$ and $k_s$ are the two arbitrary values that the multiplicity function can take on~$B_L$.
The action of $B_L$ on $(\CC^N)^{\otimes L}$ is taken as
\begin{gather*}
R_{t_i}=P_{ii+1}\qquad\text{and}\qquad R_r=Q_L,
\end{gather*}
where $Q\in {\rm End}(\CC^N)$ satisf\/ies $Q^2=\II_{N}$. We apply (\ref{maps}) to
\begin{gather*}
T_a(u)=\sum_{k=1}^L\frac{P_{ak}}{u-d_k}
\end{gather*}
with $\sigma=Q$ (so $\tau=-1$ and $n=2$).
We get
\begin{gather*}
B_a(u)=\frac{1}{2}\left(T_a(u)-Q_aT_a(-u)Q_a\right)=\frac{1}{2}\sum_{k=1}^L
\left(\frac{P_{ak}}{u-d_k}+\frac{Q_aP_{ak}Q_a}{u+d_k}\right).
\end{gather*}
To prove that $B(u)\Lambda$ satisf\/ies (\ref{twisted_rel}), it is enough to prove
\begin{gather}
\label{proof}
\hat{r}R_rB(u)=B(u)\hat{r}R_r\qquad \text{and}\qquad \hat{t_i}R_{t_i}B(u)=B(u)\hat{t_i}R_{t_i} .
\end{gather}
This is readily seen as
\begin{gather*}
\hat{r}R_rB(u)=\frac{1}{2}\sum_{k=1}^{L-1}\left(\frac{P_{ak}}{u-d_k}+\frac{Q_aP_{ak}Q_a}{u+d_k}\right)\hat{r}R_r
 +\frac{1}{2}
\left(\frac{Q_LP_{aL}Q_L}{u-\hat{r}d_L\hat{r}}+\frac{Q_LQ_aP_{aL}Q_aQ_L}{u+\hat{r}d_L\hat{r}}\right)\hat{r}R_r
\end{gather*}
and $\hat{r}d_L\hat{r}=d_{re_L}=d_{-e_L}=-d_L$ (recall also that $Q^2=\II_N$). Similarly, one looks at the $i$-th and
$i+1$-th term in the second relation in (\ref{proof}) and uses $\hat{t_i}d_i\hat{t_i}=d_{t_ie_i}=d_{e_{i+1}}=d_{i+1}$ together with $P_{ii+1}P_{ai}P_{ii+1}=P_{ai+1}$.

One can now extract the commuting quantities by expanding $b(u)\Lambda$ in powers of $u^{-1}$. One gets
\begin{gather}
b(u)\Lambda=\sum_{n=0}^\infty\frac{1+(-1)^n}{2u^{n+1}}\sum_{k=1}^Ld_k^n\Lambda .
\end{gather}
We see that only $  J_{2n}=\sum\limits_{k=1}^Ld_k^{2n}\Lambda$ yield non trivial quantities. In particular, one gets the usual~$B_L$ spin
Calogero Hamiltonian
\begin{gather}
\label{BC_Ham}
H=J_2=-\sum_{j=1}^L \partial_{e_j}^2+k_l\sum_{m\neq j}
\left(
\frac{k_l-P_{mj}}{(x_m-x_j)^2}
+\frac{k_l-Q_jP_{mj}Q_j}{(x_m+x_j)^2}
\right)
+k_s\sum_{j=1}^L\frac{k_s-Q_j}{x_j^2}.
\end{gather}

Then, we get the well-known result \cite{Ya} that this hamiltonian is integrable by proving that
$J_2,J_4,\dots,J_{2L}$ are independent. In addition, as explained in Section~\ref{sec:link}, we prove that this model has for symmetry the twisted half-loop algebra
of order~2. This result may be also obtained by
considering the symmetry of the $B_L$ Sutherland model proved in~\cite{caduk}
and taking the suitable limit as explained in~\cite{BGHP} for
the $A_L$ case.

\subsection[Another $B_L$ spin Calogero model]{Another $\boldsymbol{B_L}$ spin Calogero model} \label{another}

Here we follow the approach of \cite{ino} and use another representation of $B_L$ for
the spins while keeping the same on $\RR^L$.
Let us f\/ix $\mu \in \R^L$.
The orbit of the vector $\mu$ under the group $B_L$ is written
\begin{gather*}
\{w(\mu)\ |\ w\in B_L \}\equiv\{\mu_1,\dots,\mu_M\}
\end{gather*}
for some $M\in \NN$ and with $\mu_j\in \RR^L$. The group $B_L$ acts transitively on this set,
i.e.\  for any~$\mu_j$,~$\mu_k$ there exists $w\in B_L$ such that
$w(\mu_j)=\mu_k$.
For any $w\in B_L$, we obtain an action on the set $\{1,\dots, M\}$
by def\/ining
\begin{gather}
\label{def_check}
\check w(i)=j \qquad\text{if\/f}\qquad w(\mu_i)=\mu_j.
\end{gather}
The map $w\mapsto\check w$ is a representation of $B_L$. We can now def\/ine a representation of $B_L$ on
$(\mathbb{C}^N)^{\otimes M}$ as the map $w\mapsto R_w$ where
\begin{gather}
\label{spin_rep}
R_w v_1\otimes\dots\otimes v_M=v_{\check w^{-1}(1)}
\otimes\dots\otimes v_{\check w^{-1}(M)}
\end{gather}
with $v_i\in \mathbb{C}^N$.

At this stage, one can see that
\begin{gather*}
T_a(u)=\sum_{k=1}^{M}\frac{P_{ak}}{u-d_{\mu_k}}
\end{gather*}
commutes with the projector $\Lambda$ in the chosen representation. Indeed, for each $w\in B_L$
\begin{gather*}
\hat{w}R_w \sum_{k=1}^{M}\frac{P_{ak}}{u-d_{\mu_k}}=\sum_{k=1}^{M}\frac{P_{a\check{w}(k)}}{u-d_{w(\mu_k)}}\,\hat{w}R_w=
\sum_{n=1}^{M}\frac{P_{a n}}{u-d_{\mu_n}}\,\hat{w}R_w  ,
\end{gather*}
where the last equality is obtained by relabelling the sum according to (\ref{def_check}).
Thus, $T(u)\Lambda$ satisf\/ies the half-loop algebra relations and ${\rm Tr}\,T(u)\Lambda\equiv
t(u)\Lambda$ provides the commuting elements.

So we f\/ind that all the $B_L$ spin Calogero Hamiltonians obtained in this way (e.g.~in~\cite{ino})
have the half-loop algebra as symmetry algebra.

As an example, we apply this to $\mu=e_1$, the f\/irst vector of the canonical basis of $\RR^L$. Then the orbit contains $2L$ elements $\pm e_j$, $j=1,\dots,L$
which we order as follows $(e_1,-e_1,\dots,e_L,-e_L)\equiv(\mu_1,\mu_{\bar 1},\dots,\mu_{L},\mu_{\bar L})$. Then, by inspection we get
\begin{gather*}
\check r(j)=\begin{cases}
\bar L,~~& j=L,\\
L,~~& j=\bar L ,\\
j,~~& \text{otherwise}
\end{cases}
\qquad \text{and}\qquad
\check t_i(j)=\begin{cases}
i+1,&~~j=i ,\\
i,&~~j=i+1 ,\\
\overline{i+1},&~~j=\bar i ,\\
\bar i,&~~j=\overline{i+1},\\
j,&~~\text{otherwise}.
\end{cases}
\end{gather*}
This gives
\begin{gather*}
R_r=P_{L,\bar L} ,\qquad R_{t_i}=P_{i,i+1}P_{\bar i,\overline{i+1}}.
\end{gather*}
We get
\begin{gather*}
T_a(u)=\sum_{k=1}^{L}\frac{P_{ak}}{u-d_{\mu_k}}+\frac{P_{a\bar k}}{u-d_{\mu_{\bar k}}}
=\sum_{k=1}^{L}\frac{P_{ak}}{u-d_{k}}+\frac{P_{a\bar k}}{u+d_{k}},
\end{gather*}
where the Dunkl operators are given by (\ref{def2_dunkl}).
The following $B_L$ spin Calogero Hamiltonian has the half-loop algebra as symmetry
algebra
\begin{gather}
\label{another_Ham}
H=-\sum_{j=1}^L \partial_{e_j}^2+k_l\sum_{m\neq j}
\left(
\frac{k_l-P_{mj}P_{\bar m\bar j}}{(x_m-x_j)^2}
+\frac{k_l-P_{m\bar j}P_{\bar m j}}{(x_m+x_j)^2}
\right)+k_s\sum_{j=1}^L\frac{k_s-P_{j\bar j}}{x_j^2}.
\end{gather}
As previously, ${\rm Tr}\, T(u)\Lambda$ provides only the even conserved quantities
$J_{2k}=\sum\limits_{j=1}^L d_j^{2k}\Lambda$ but it is suf\/f\/icient to prove the integrability
since $J_2, J_4,\dots,J_{2L}$ are again independent. Comparing (\ref{BC_Ham}) and (\ref{another_Ham}), it is manifest that the only
dif\/ference lies in the action on the spins, the potentials being indentical. Yet, these two systems based on $B_L$ have dif\/ferent symmetry algebras.
It is also interesting to remark that (\ref{another_Ham}) represents a system of $L$ particles on the line
with two spin degrees of freedom attached to each particle.

\section[Group $G_2$]{Group $\boldsymbol{G_2}$}\label{G2}

We consider the Coxeter group $G_2$
(which is the dihedral group $I_2(6)$ of order 12). It is generated by 2
elements $t$ and $r$ subject only to the following relations
\begin{gather*}
r^2=1 ,\qquad t^2=1  ,\qquad (tr)^6=1 .
\end{gather*}
It may also be generated by 2 elements $a$ and $b$ subject only to the following relations
\begin{gather*}
a^6=1  ,\qquad b^2=1  ,\qquad ba=a^{-1}b .
\end{gather*}
The isomorphism between the two presentations reads
\begin{gather*}
a\mapsto tr\qquad\text{and}\qquad b\mapsto t.
\end{gather*}

\subsection{Model with three particles and six spins}\label{sec:I1}

The choice in this section for the positive roots of $I_2(6)$ in $\RR^3$ are
$\Phi^+=\{e_1-e_2,e_3-e_1,e_3-e_2\}\cup\{-2e_1+e_2+e_3,e_1-2e_2+e_3,-e_1-e_2+2e_3\}$.
Then, the the representation of the generators of $I_2(6)$ associated to
simple roots are
\begin{gather}
\label{eq:ac3}
t=s_{e_1-e_2}=\left(\begin{array}{c c c}
0 & 1 & 0\\
1 & 0 & 0\\
0 & 0 & 1
\end{array}\right)\qquad\text{and}\qquad
r=s_{-2e_1+e_2+e_3}=\frac{1}{3}\left(\begin{array}{c c c}
-1 & 2 & 2\\
2 & 2 & -1\\
2 & -1 & 2
\end{array}\right).
\end{gather}
To write down the Dunkl operators explicitly, it is useful to express
all the ref\/lections in terms of $r$ and $t$
\begin{gather*}
s_{e_1-2e_2+e_3}=trt  ,\qquad
s_{e_1-e_3}=rtr , \qquad
s_{e_2-e_3}=trtrt  ,\qquad
s_{e_1+e_2-2e_3}=rtrtr.
\end{gather*}
We give only one example as the other ones are computed similarly:
\begin{gather*}
d_{e_1}=-i\partial_{e_1}+ik_s\left(\frac{1}{x_1-x_2}\hat t
+\frac{1}{x_1-x_3}\widehat{rtr}\right)
\nonumber\\
\phantom{d_{e_1}=}{}
+ik_l\left(\frac{1}{x_1-2x_2+x_3}\widehat{trt}
-2\frac{1}{-2x_1+x_2+x_3}\hat r
+\frac{1}{x_1+x_2-2x_3}\widehat{rtrtr}\right).
\end{gather*}

The action on the spins is def\/ined as in Section~\ref{another} replacing the group $B_L$ by $I_2(6)$.
That is, we def\/ine for $\mu \in \R^3$,
\begin{gather*}
\{\mu_1,\dots,\mu_M\}=\{w(\mu)\ | \ w\in I_2(6)\}
\end{gather*}
which provides a representation $R_w$ acting on spins.
Thus, we have again
that
\begin{gather*}
T_a(u)=\sum_{k=1}^{M}\frac{P_{ak}}{u-d_{\mu_k}}
\end{gather*}
commutes with the projector $\Lambda$ in the chosen representation
(the proof follows the same lines than the one in Section~\ref{another}).
We conclude that the half-loop algebra is the symmetry algebra
of the integrable hierarchy contained in ${\rm Tr}\,T(u)\Lambda$.

We give an example. We may choose $\mu=e_1$. Then the orbit is
\begin{gather*}
\Big\{\mu_1=e_1, \ \mu_2=e_2,\ \mu_3=e_3, \
\mu_{\bar 1}=\tfrac{1}{3}(-e_1+2e_2+2e_3), \nonumber\\
\qquad
\mu_{\bar 2}=\tfrac{1}{3}(2e_1-e_2+2e_3), \ \mu_{\bar 3}=\tfrac{1}{3}(2e_1+2e_2-e_3)\Big\}.
\end{gather*}
So, we get $M=6$.
We deduce the following representation of $I_2(6)$ on $(\CC^N)^{\otimes 6}$
\begin{gather*}
R_t=P_{12}P_{\bar 1 \bar 2}\qquad\text{and}\qquad R_r=P_{1\bar 1}P_{2\bar 3}P_{3\bar 2}.
\end{gather*}
The coef\/f\/icient in front of $u^{-2}$ in the expansion of $t(u)\Lambda$ is proportional to the total momentum
$P=-i\partial_1-i\partial_2-i\partial_3$ and the one in front of $u^{-3}$ is proportional to the Hamiltonian
and reads
\begin{gather*}
H=-\sum_{i=1}^3 \partial_{e_i}^2 +k_s\sum_{m\neq j}\frac{k_s-P_{mj}P_{\bar m\bar j}}{(x_m-x_j)^2}
+k_l\sum_{n\neq m\neq j}\frac{k_l-P_{n\bar n}P_{j\bar m}P_{m\bar j}}{(-2x_n+x_m+x_j)^2}.
\end{gather*}
By direct computation, we can show that the coef\/f\/icients in front of $u^{-4},\ u^{-5}$ and $u^{-6}$ are not
independent of $P$ and $H$ whereas $J_{6}$, the one in front of $u^{-7}$, is independent of them.
Then $P$, $H$ and $J_6$ provides 3 independent conserved quantites which proves the integrability of~$H$.
Let us remark that the degree in $d_i$ of the conserved quantities (i.e.\  1,~2,~6) are in agreement with the
degrees of the invariant polynomials by the Coxeter group $I_2(6)$. In fact, it is not a~coincidence since
the conserved quantities we constructed may be seen as polynomials in terms of the variables $d_i$ and are,
by construction, invariant under $I_2(6)$.

\subsection{Model with three particles and three spins}\label{sec:I2}

We keep the same positive roots as in Section~\ref{sec:I1}. Then, the action of $I_2(6)$ on $\RR^3$ is still
the one def\/ined by (\ref{eq:ac3}) but we take now the following representation of $I_2(6)$ on
$(\CC^N)^{\otimes 3}$
\begin{gather*}
R_{r}=P_{23}Q_1Q_2Q_3 ,\qquad R_{t}=P_{12},
\end{gather*}
where $Q^2=\II_{N}$.
The projector may be written as follows
\begin{gather*}
\Lambda = \Lambda_Q\Lambda_P=\Lambda_P\Lambda_Q,\\
\Lambda_Q = \tfrac{1}{2} (1+(\widehat{tr})^3\, Q_1Q_2Q_3),\\
\Lambda_P =\tfrac{1}{6}(1+\widehat{t}\ P_{12}
+\widehat{trtrt}\, P_{23}
+\widehat{rtrt}\, P_{12}P_{23}
+\widehat{trtr}\, P_{23}P_{12}
+\widehat{rtr}\, P_{13}).
\end{gather*}
The projector $\Lambda_P$ is the usual symmetriser of the eigenfunctions for bosons
and $\Lambda_Q$ yields
\begin{gather*}
\phi(x_1,x_2,x_3\ |\ s_1,s_2,s_3)=
\phi(-x_1+2X,-x_2+2X,-x_3+2X\ |\ s_1^*,s_2^*,s_3^*),
\end{gather*}
where $s^*=Qs$ and $X=\frac{1}{3}(x_1+x_2+x_3)$ is the center of mass of the three particles.

The monodromy matrix is taken to be (with $d_i\equiv d_{e_i}$),
\begin{gather*}
T_a(u)=\frac{P_{a1}}{u-d_1}+\frac{P_{a2}}{u-d_2}+\frac{P_{a3}}{u-d_3}.
\end{gather*}
We apply a slight variant of (\ref{maps}) and def\/ine
\begin{gather*}
\hat B_a(u) = T_a(u)-Q_aT_a(-u+D)Q_a\nonumber\\
\phantom{\hat B_a(u)}{}= \frac{P_{a1}}{u-d_1}+\frac{P_{a2}}{u-d_2}+\frac{P_{a3}}{u-d_3}+
\frac{Q_1P_{a1}Q_1}{u+d_1-D}+\frac{Q_2P_{a2}Q_2}{u+d_2-D}+\frac{Q_3P_{a3}Q_3}{u+d_3-D},
\end{gather*}
where $D=\frac{2}{3}(d_1+d_2+d_3)$. $\hat B(u)$ satisf\/ies the following commutation relations
\begin{gather}
\label{commute_D}
\big[\hat B_a(u),\hat B_b(v)\big]=\frac{1}{2}\left[\hat B_a(u)+\hat B_b(v),
\frac{P_{ab}}{u-v}\right]+
\half\left[-\hat B_a(u)+\hat B_b(v),
\frac{Q_aP_{ab}Q_a}{u+v-D}\right].
\end{gather}
This algebra is isomorphic to the twisted half-loop algebra of order $2$: $\hat B(u)\mapsto B(u+D/2)$.
The crucial points now are that $\hat B(u)$ commutes with $\Lambda$ and we still have
$[{\rm Tr}\, \hat B(u),{\rm Tr}\, \hat B(v)]=0$.
Then, the coef\/f\/icient of $u^{-2}$ in $b(u)\Lambda$ is proportional to the total momentum
$P=-i\partial_1-i\partial_2-i\partial_3$
and the one of $u^{-3}$ contains the following Hamiltonian, studied in~\cite{CQ},
\begin{gather}
\label{eq:ham2}
H=d_1^2+d_2^2+d_3^2=-\sum_{i=1}^3 \partial_{e_i}^2 +k_s\sum_{m\neq j}\frac{k_s-P_{mj}}{(x_m-x_j)^2}
+k_l\sum_{n\neq m\neq j}\frac{k_l-P_{mj}Q_nQ_mQ_j}{(-2x_n+x_m+x_j)^2}.
\end{gather}
As before, the coef\/f\/icients of $u^{-4}$, $u^{-5}$ and $u^{-6}$ are not independent of~$P$ and~$H$.
It is again~$J_6$, the coef\/f\/icient of $u^{-7}$, which provides the third independent conserved quantity and allows us to
prove that the previous Hamiltonian is integrable.

Concerning the symmetry algebra, one has to be careful since the commutation
relations~(\ref{commute_D}) depend on the operator $D$ which is proportional to the total momentum.
We can talk about the symmetry of the model (\ref{eq:ham2}) only in the sectors where the total momentum $P$ has a given value~$p$.
This is not a problem as the states are def\/ined by three quantum numbers, $p$ being one of them and in the sectors of f\/ixed $p$,
the symmetry algebra is isomorphic to the twisted half-loop algebra.
For example, in the center of the mass frame ($p=0$), the symmetry algebra of (\ref{eq:ham2}) is the
usual twisted half-loop algebra of order $2$.

\subsection{Model with two particles and two spins}\label{sec:I3}

Our choice for the simple roots of $I_2(6)$ in $\RR^2$ is
$e_1-e_2$ and $-e_1+\cot\left(\frac{\pi}{12}\right)e_2$.
Then the other 4 positive roots are
\begin{gather*}
e_1-\tan\left(\frac{\pi}{12}\right)e_2 , \qquad
e_1+\tan\left(\frac{\pi}{12}\right)e_2  , \qquad
e_1+e_2 , \qquad
e_1+\cot\left(\frac{\pi}{12}\right)e_2.
\end{gather*}
Then, the action of both generators of $I_2(6)$ on $\RR^2$ is given explicitly by
\begin{gather*}
b=\left(\begin{array}{c c}
0 &1\\
1& 0
\end{array}\right)\qquad\text{and}\qquad\
a=
\left(\begin{array}{c c}
\cos(\frac{\pi}{3})&-\sin(\frac{\pi}{3}) \vspace{1mm}\\
\sin(\frac{\pi}{3})&\cos(\frac{\pi}{3})
\end{array}\right).
\end{gather*}
It is useful to express all the ref\/lections in terms of $a$ and $b$
\begin{gather*}
 s_{e_1-e_2}=b  ,\qquad
s_{e_1-\tan\left(\frac{\pi}{12}\right)e_2}=ab ,\qquad
s_{e_1+\tan\left(\frac{\pi}{12}\right)e_2}=a^2b ,\\
 s_{e_1+e_2}=ba^3=a^3b ,\qquad
s_{e_1+\cot\left(\frac{\pi}{12}\right)e_2}=ba^2 ,\qquad
s_{-e_1+\cot\left(\frac{\pi}{12}\right)e_2}=ba  .
\end{gather*}
Then, for example, we get
\begin{gather*}
d_{e_1}=-i\partial_{e_1}+
ik_s\left(
\frac{1}{x_1-x_2}\hat b+\frac{1}{x_1+(2-\sqrt{3})x_2}\hat a^2 \hat b
+\frac{1}{x_1+(2+\sqrt{3})x_2}\hat b\hat a^2\right)\nonumber\\
\phantom{d_{e_1}=}{} +ik_l\left(\frac{1}{x_1-(2-\sqrt{3})x_2}\hat a\hat b
+\frac{1}{x_1+x_2}\hat a^3b+
\frac{1}{x_1-(2+\sqrt{3})x_2}\hat b\hat a\right),
\end{gather*}
where we have used $\tan(\frac{\pi}{12})=2-\sqrt{3}$ and
$\cot(\frac{\pi}{12})=2+\sqrt{3}$.
We def\/ine the action on $(\CC^N)^{\otimes 2}$ by
\begin{gather*}
  R_{a}=Q_1Q_2^{-1}  ,\qquad R_{b}=P_{12},
\end{gather*}
where $Q^6=\II_{N}$.
The projector $\Lambda$ may be written as follows
\begin{gather}
\label{eq:pro5}
\Lambda=\Lambda_Q\Lambda_P=\Lambda_P\Lambda_Q,\\
\Lambda_P=\half(1+\hat b\ P_{12}),\nonumber\\
\Lambda_Q=\frac{1}{6}\Big(1+\hat{a} Q_{1}Q_{2}^{-1}
+\hat{a}^2 (Q_1Q_2^{-1})^{2}
+\hat{a}^3 (Q_2Q_1^{-1})^{3}
+\hat{a}^4 (Q_1Q_2^{-1})^{4}
+\hat{a}^5 (Q_2Q_1^{-1})^{5}\Big).\nonumber
\end{gather}
So, it is easy to see that the eigenfunctions are totally symmetric
$\phi(x_1,x_2 | s_1,s_2){=}\phi(x_2,x_1 | s_2,s_1)$ and satisfy the additional relation
\begin{gather*}
\phi(x_1,x_2 \, | \, s_1,s_2)=\phi\left(\frac{x_1+\sqrt{3}x_2}{2},\frac{x_2-\sqrt{3}x_2}{2}
\, \Big| Qs_1,Q^{-1}s_2\right).
\end{gather*}
It turns out to be convenient to map $\RR^2$ into $\CC$.
Let us def\/ine
\begin{gather*}
d=d_{e_1}+id_{e_2}\qquad\text{and}\qquad\overline{d}=d_{e_1}-id_{e_2} ,
\end{gather*}
and similarly for $\partial$, $\bar \partial$.
We write the Dunkl operators in terms of $z=x_1+ix_2$ and $\overline z=x_1-ix_2$
\begin{gather*}
d=-i\partial+
2\sum_{j=0}^2\left(
\frac{ik_s}{i\tau^{2j}z+\overline z}\hat b\hat a^{2j}
+\frac{ik_l}{i\tau^{2j+1} z+\overline z}\hat b\hat a^{2j+1}\right),\\
\overline d = -i\bar\partial+
2\sum_{j=0}^2\left(
\frac{ik_s}{z-i\tau^{2j}\overline z}\hat a^{2j}\hat b
+\frac{ik_l}{z-i\tau^{2j+1}\overline z}\hat a^{2j+1}\hat b\right),
\end{gather*}
where $\tau=\exp(i\pi/3)$. The action of the generators of $I_2(6)$ on these Dunkl operators are
\begin{gather*}
\hat ad=\tau^{-1} d\hat a,\qquad \hat a\overline d=\tau \overline d\hat a,\qquad
\hat bd=i\overline d \hat b
\qquad\text{and}\qquad  \hat b\overline d=-i d \hat b .
\end{gather*}
Now, we introduce
\begin{gather*}
T_a(u)=\frac{P_{a1}}{{u-d}}+\frac{P_{a2}}{u-i{\overline d}}
\end{gather*}
and also
\begin{gather*}
\hat B_a(u)=\sum_{j=0}^5 \tau^j Q_a^{-j} T_a((u-d\overline d)\tau^j) Q_a^{j}
=\sum_{j=0}^5 \frac{Q_a^{-j}P_{a1}Q_a^{j}}{u-d\overline d-\tau^{-j}d}
+\frac{Q_a^{-j}P_{a2}Q_a^{j}}{u-d\overline d-i\tau^{-j}{\overline d}} .
\end{gather*}
This satisf\/ies the commutation relations of the twisted half-loop algebra of order 6
with a shift of $d\overline d$ in the spectral parameters. We can show by direct investigation that
$\hat B(u)$ commutes with the projector def\/ined by~(\ref{eq:pro5}).
As usual, the series ${\rm Tr}\, \hat B(u)\Lambda$ provides the conserved quantities.
We can show that 2 are independent: the coef\/f\/icients in front of $u^{-2}$ and $u^{-7}$
proportional respectively to
\begin{gather*}
H=d\, \overline d\qquad\text{and}\qquad J_6=d^6-\overline d^6+2(d\, \overline d)^6 .
\end{gather*}
Explicitly, the Hamiltonian is
\begin{gather*}
H=-\partial_z\partial_{\overline{z}}
+4i\sum_{j=0}^2\left(
k_s\tau^{2j}\frac{Q_1^{-2j}P_{12}Q_1^{2j}-k_s}{(z-i\tau^{2j}\overline z)^2}
+k_l\tau^{2j+1}\frac{Q_1^{-(2j+1)}P_{12}Q_1^{2j+1}-k_l}{(z-i\tau^{2j+1}\overline z)^2}\right).
\end{gather*}
As in Section~\ref{sec:I2}, the commutation relations satisf\/ied by $\hat B(u)$ depend on one
operator of our theory ($H$). Thus, in each sector of f\/ixed energy, the symmetry algebra of the integrable hierarchy is
isomorphic to the twisted half-loop algebra of order $6$.

\section{Generalizations}\label{general}

Let us remark that the construction of Section~\ref{sec:I3} for $I_2(6)$ extends to $I_2(m)$, a presentation of which is
\begin{gather*}
a^m=1 ,\qquad b^2=1  ,\qquad ba=a^{-1}b .
\end{gather*}
The results follow by taking $Q$ with $Q^m=\II_N$ and $\tau=e^{\frac{2i\pi}{m}}$. In the expressions for the Dunkl, one has to modify the sum
and the two independent operators are to be found at orders~$u^{3}$ and~$u^{m+1}$.
The relevant symmetry algebra is the twisted half-loop algebra of order $m$.

Another general result holds: the construction of the spin representation described in Sections~\ref{another} and \ref{sec:I1} applies to any f\/inite
Coxeter group $W$ as explained in~\cite{ino}. So f\/ixing $\mu$ and denoting $\{\mu_1,\dots,\mu_M\}$ the orbit of $\mu$ under $W$, the representation $R$
as def\/ined in (\ref{spin_rep}) is always well-def\/ined on $(\CC^N)^{\otimes M}$ and the monodromy matrix
\begin{gather*}
T_a(u)=\sum_{k=1}^{M}\frac{P_{ak}}{u-d_{\mu_k}}
\end{gather*}
commutes with the projector $\Lambda$ in the chosen representation.
So we f\/ind that all the $W$ spin Calogero Hamiltonians obtained in this way have the half-loop algebra as symmetry algebra.

At this stage, it is worth mentioning that potentially physically interesting integrable Hamiltonians can always be constructed. Indeed, from the
general theory of invariant polynomials under f\/inite Coxeter groups, the polynomial of order two in the variables is always invariant. This means that
Hamiltonians of the form
\begin{gather*}
-\sum_{j=1}^L\partial_{e_i}^2+\text{interactions terms}
\end{gather*}
can always be extracted from the hierarchy. We note however that $M$ can be dif\/ferent from $L$ so these systems may represent particles with more than one
spin attached to them.

\section{Conclusions}

We have shown that, as far as symmetry is concerned, speaking of a $W$ spin Calogero model is not enough. Even if the root structure of $W$
provides the form of the interaction term in Hamiltonians constructed from Dunkl operators based on $W$, the symmetry algebra heavily depends on
the choice of representation of $W$ on the spins. Using the method pioneered in \cite{BGHP}, we showed this explicitly for the case of the Coxeter groups
$B_L$ and $I_2(6)$. For the former, two dif\/ferent symmetry algebras were constructed while we proposed three of them for the latter.
We stress that in all cases, the identif\/ication of the symmetry algebra is an original result in itself.

It should be noted
that one systematic method, based on \cite{ino}, always yields the same symmetry algebra (here the half-loop algebra for Calogero type models) for all
Coxeter groups. However,
there seems to be no systematic understanding of how to produce other possibilities such as those of Section~\ref{sec:1} for $B_L$ and Sections~\ref{sec:I2},
\ref{sec:I3} for $G_2$. In physical terms, such an understanding would allow one to control the number of spins $M$ and make it coincide (if desired)
with the number of particles $L$. Indeed, it all amounts to f\/inding a construction of spin representations based on orbits of conveniently chosen subgroups
of~$W$. We hope to come back to this question in the future.

Other directions of investigation could involve the so-called Sutherland models, whose traditional study involves the same tools as described in this paper.
However, it is expected that novel algebraic structures should be understood f\/irst. Indeed, already in the $A_L$ case, the symmetry algebra for spin Sutherland
models is related to the Yangian of $\mathfrak{gl}_N$, a deformation of the half-loop algebra. For the $B_L$ case, the ref\/lection algebra appears (when choosing the
same spin representation as in Section~\ref{sec:1}) which is a deformation of the twisted half-loop algebra of order~$2$. However, for higher order twisting,
the corresponding deformations are not known.

Finally, the symmetry algebras of the dif\/ferent spin Calogero models obtained in this paper might be useful to compute their eigenstates.
Indeed, in the case of the $A_L$ Sutherland model, the Yangian symmetry was the cornerstone of the explicit construction of the eigenvectors~\cite{TU,Tak}.
It is based on representation theory of the Yangian and, in particular, on the construction of the Gelfand--Zetlin bases~\cite{NT} which uses the knowledge of the
 maximal Abelian subalgebra containing the center. Let us remark that strangely enough, the situation in the case of the half-loop algebra seems more complicated.
 Indeed, the maximal Abelian subalgebra of the half-loop algebra has been discovered 10 years after the one of the Yangian~\cite{Tala}.
 Thus, we hope that the present paper provides one additional motivation to look for the maximal Abelian subalgebra of the twisted cases.

\subsection*{Acknowledgements}
N.C. would like to thank the hospitality of the Centre for Mathematical Science, City University, where this work was initiated.

\pdfbookmark[1]{References}{ref}
\LastPageEnding
\end{document}